\documentclass[twocolumn,secnumarabic,amssymb, nobibnotes, aps, prd]{revtex4}

\usepackage{amsmath,amssymb}
\usepackage{graphicx}
\usepackage{setspace}
\usepackage[sort&compress]{natbib}
\usepackage{hyperref}
\hypersetup{colorlinks=true, citecolor=blue}

\begin{document}
\title{Growth of cosmic perturbations in the modified $f(R,T)$ gravity}

\author{Mahnaz Asghari}
\email{mahnaz.asghari@shirazu.ac.ir}
\affiliation{Department of Physics, College of Science, 
	Shiraz University, Shiraz 71454, Iran \\
	Biruni Observatory, College of Science, 
	Shiraz University, Shiraz 71454, Iran} 
\author{Ahmad Sheykhi}
\email{asheykhi@shirazu.ac.ir}
\affiliation{Department of Physics, College of Science, 
	Shiraz University, Shiraz 71454, Iran \\
	Biruni Observatory, College of Science, 
	Shiraz University, Shiraz 71454, Iran} 

\begin{abstract}
We explore the generalized $f(R,T)$ modified theory of gravity,
where the gravitational Lagrangian is a function of Ricci scalar
$R$ and the trace of the energy-momentum tensor $T$. We derive
modified field equations to the linear order of perturbations in
the context of $f(R,T)$ model. We then investigate the growth of
perturbations in the context of $f(R,T)$ modified gravity. Primary
numerical investigations based on matter power spectra diagrams
indicate a structure growth suppression in $f(R,T)$ gravity,
which exhibits consistency with local measurements. Also, we
notice that matter-geometry interaction in $f(R,T)$ model would
results in the specific feature named as "matter acoustic
oscillations" appeared in matter power spectra diagrams. Moreover,
we put constraints on the cosmological parameters of $f(R,T)$
model, utilizing current observations, chiefly cosmic microwave
background (CMB), weak lensing, supernovae, baryon acoustic
oscillations, and redshift-space distortions data.
Numerical results based on MCMC calculations imply that $f(R,T)$
is a qualified theory of modified gravity in reconciling Planck
CMB data with local probes of large scale structures, by reporting
lower values for the structure growth parameter $\sigma_8$
compared to the standard model of cosmology.
\end{abstract}

\maketitle

\section{Introduction}
It is a general belief that Einstein's theory of general
relativity (GR) is one of the most comprehensive theories of
gravity which has proven to be highly efficient in understanding
the nature of gravity as well as fundamental concepts in
cosmology. Furthermore, the discovery of acceleration of the
cosmic expansion which is supported by observational measurements
from type Ia supernovae (SNeIa) \cite{sn1,sn2}, affirms the
concordance $\Lambda$CDM model in the framework of GR, as a well
established cosmological model for describing the evolution of the
universe. Although the standard model of cosmology is successfully
verified by a broad range of observations containing the cosmic
microwave background (CMB) anisotropies \cite{cmb1,cmb2,cmb3},
large scale structures (LSS) \cite{lss1,lss2,lss3}, and baryon
acoustic oscillations (BAO) \cite{bao1,bao2,bao3}, this model
suffers from several theoretical and observational shortcomings.
In principle, low redshift cosmological probes prefer less
structure formation which yields lower values of linear matter
density perturbation amplitude $\sigma_8$, compared to the global
predictions based on the Planck CMB data
\cite{s81,s82,s83,s84,s85,s86}. Moreover, the inferred values of
Hubble constant according to local determinations are in conflict
with CMB measurements \cite{H01,H02,H03,H04,H05,H06}. In this
regard, it seems that GR is not necessarily an ultimate gravity
theory, and thus there is a tendency towards alternative theories
of gravity.

The modified gravity theory which is based upon corrections
to the Einstein-Hilbert (EH) action, is introduced as a
convincing alternative model of gravity, in purpose of
describing the emerged issues in GR where the standard model
of cosmology is not capable of solving them.
The most straightforward modification of gravity is to
consider an arbitrary function of the Ricci scalar $R$
in the EH action, known as $f(R)$ gravity
\cite{fr1,fr2,fr3,fr4,fr5,fr6,fr7,fr8,fr9,fr10}.
It is also convenient to consider a non-minimal coupling
between matter and geometry, as firstly proposed in
\cite{frlm1,frlm2}, where an explicit coupling between
the Ricci scalar $R$ and the matter Lagrangian density $L_m$
is contemplated. In this generalization of modified gravity,
called $f(R,L_m)$ theory, the equation of motion
of massive particle is non-geodesic, and thus an extra
force arises. The $f(R,L_m)$ gravity has been extensively
explored in the literature
\cite{frlm3,frlm4,frlm5,frlm6,frlm7,frlm8,frlm9,frlm10,
     frlm11,frlm12}.
Accordingly, based on the non-minimal curvature-matter
interaction, Harko et al. \cite{frt1} introduced the
generalized gravity model $f(R,T)$, in which the gravitational
Lagrangian is a function of Ricci scalar $R$ and the trace
of the energy-momentum tensor $T$.
The $T$-dependent Lagrangian in this class of modifications
on GR, indicates quantum effects which results in a
non-conservative energy-momentum tensor. Thus, an extra
acceleration would appear induced by the coupling between
matter and geometry \cite{thermo1,thermo2}.
There are several related studies in the framework
of $f(R,T)$ gravity, where we specify some of them in
the following. Jamil et al. \cite{frt2} considered
reconstruction of different cosmological models
in $f(R,T)$ gravity. Cosmological solutions
of $f(R,T)$ theory via phase space analysis was explored by
Shabani and Farhoudi \cite{frt3}. Sharif and Nawazish
\cite{frt4} examined the existence of Noether symmetry
in the context of $f(R,T)$ modified gravity.
A discussion on energy conditions applications as well as the
cosmological viability of $f(R,T)$ gravity was provided by
Moraes and Sahoo \cite{frt5}. Rajabi and Nozari \cite{frt6}
investigated inflation and reheating era in the unimodular
$f(R,T)$ theory of gravity. Moreover, Fortunato et al. \cite{frt7}
reconstructed modified $f(R,T)$ model through Gaussian process.
For more related investigations on $f(R,T)$ gravity refer to
\cite{frt8,frt9,frt10,frt11,frt12,frt13,frt14,
    frt15,frt16,frt17,frt18,frt19,frt20,frt21,
    frt22,frt23,frt24,frt25,frt26,frt27,frt28,
    frt29,frt30,frt31,frt32,frt33,frt34,frt35}.
On the other hand, the parameterized post-Newtonian solar system
analysis put severe constraints on $f(R,T)$ gravity with linear
dependence on $T$ \cite{frtss}, where can not rule out this class of
modified gravity models on cosmological scales.
Then, it is worth turning our attention to put observational constraints on $f(R,T)$ model via numerical analysis.
While cosmological background data are vastly utilized as
observational probes
\cite{frtobs1,frtobs2,frtobs3,frtobs4,frtobs5,frtobs6,frtobs7,frtobs8,frtobs9},
it is pertinent to take into account cluster counts
measurements along with background observations in an attempt to
explore cosmic structure growth in modified $f(R,T)$ theory.
In this regard, recently \cite{frtfs} have considered the evolution of
$f\sigma_8$ (where $f$ is the growth rate of matter perturbations)
in the context of $f(R,T)$ gravity with polynomial and
exponential $T$-dependent Lagrangian, in which severe constraints
on cosmological parameters were obtained.
Correspondingly, the purpose of the present paper is to study
$f(R,T)$ gravity in background as well as perturbation levels,
along with confronting the $f(R,T)$ model with observational
data -regarding CMB, weak lensing, supernovae, BAO, and
redshift-space distortions (RSD) measurements-
in order to provide robust constraints on cosmological parameters,
together with evaluating its ability to address the
cosmological tensions.

The plan of this paper is as follows. In section \ref{sec2} we
derive the modified field equations in $f(R,T)$ gravity. Numerical
analysis based on the $f(R,T)$ model is explained in section
\ref{sec3}. We present observational constraints on $f(R,T)$
modified gravity in section \ref{sec4}. Finally, section
\ref{sec5} provides a conclusion of our main results.
\section{Fundamental formulation of $f(R,T)$ gravity} \label{sec2}
In this section we explore the modified gravitational field
equations based on $f(R,T)$ gravity. The total action
corresponding to $f(R,T)$ modified gravity can be provided as
\cite{frt1}
\begin{align} \label{eq1}
S=\frac{1}{16\pi G}\int{\mathrm{d}^{4}x\,\sqrt{-g}f(R,T)}+\int{\mathrm{d}^{4}x\,\sqrt{-g} L_m} \,,
\end{align}
where $R$ is the Ricci scalar, $T$ denotes the trace of
the energy-momentum tensor ($T=g^{\mu \nu}T_{\mu \nu}$),
and $L_m$ stands for the matter Lagrangian density.
Contemplating action (\ref{eq1}), it is possible to derive
the associated field equations of $f(R,T)$ model as \cite{frt1}
\begin{align} \label{eq2_1}
&R_{\mu \nu}\frac{\partial f}{\partial R}-\nabla_{\mu} \nabla_{\nu}\frac{\partial f}{\partial R}
+g_{\mu \nu}\Box\frac{\partial f}{\partial R}-\frac{1}{2}fg_{\mu \nu} \nonumber \\
&=8\pi G T_{\mu \nu}-\frac{\partial f}{\partial T}\big(T_{\mu \nu}+\Theta_{\mu \nu}\big) \,,
\end{align}
with \cite{frt1}
\begin{align} \label{eq3}
\Theta_{\mu \nu}&\equiv g^{\alpha\beta}\frac{\delta T_{\alpha\beta}}{\delta g^{\mu \nu}} \nonumber \\
&=-2T_{\mu \nu}+g_{\mu \nu}L_m-2g^{\alpha\beta}\frac{\partial^2 L_m}{\partial g^{\mu \nu}\partial g^{\alpha\beta}} \,.
\end{align}
We consider the energy content of the universe as a perfect fluid
with the following energy-momentum tensor
\begin{equation} \label{eq4}
T_{\mu\nu}=\big(\rho+p\big)u_{\mu}u_{\nu}+pg_{\mu \nu} \,,
\end{equation}
where $\rho$, $p$, and $u_{\mu}$ are energy density, pressure, and
four-velocity, respectively.
It should be noted that, the on-shell matter Lagrangian of a
perfect fluid can be considered as $L_m=p$, $L_m=-\rho$,
and $L_m=T$ \cite{L1}, where the case $L_m=T$ is appropriate
for describing fluids with $0\leq w \leq 1/3$,
while $L_m=p$ and $L_m=-\rho$ are suitable for
describing dark energy fluid (in which $w\simeq-1$).
Accordingly, in the present study we choose $L_m=-\rho$,
which results in
\begin{equation} \label{eq5}
\Theta_{\mu \nu}=-2T_{\mu\nu}-\rho g_{\mu \nu} \,,
\end{equation}
where we have assumed linear dependency of $L_m$ on the metric,
being valid for this choice of matter Lagrangian \cite{L2}.
Thus, modified field equations take the form
\begin{align} \label{eq2_2}
&R_{\mu \nu}\frac{\partial f}{\partial R}-\nabla_{\mu} \nabla_{\nu}\frac{\partial f}{\partial R}+g_{\mu \nu}\Box\frac{\partial f}{\partial R}-\frac{1}{2}fg_{\mu \nu} \nonumber \\
&=8\pi G T_{\mu \nu}+\frac{\partial f}{\partial T}\big(T_{\mu \nu}+\rho g_{\mu \nu}\big) \,.
\end{align}

In the present study, for sake of simplicity, we contemplate
the following functional form of $f(R,T)$ \cite{frt1}
\begin{equation} \label{eq6}
f(R,T)=R+2f(T) \,,
\end{equation}
where
\begin{equation} \label{eq7}
f(T)=8\pi G\lambda T \,,
\end{equation}
with dimensionless constant $\lambda$. Then, modified field
equations of $f(R,T)$ theory can be written as
\begin{align} \label{eq8}
R_{\mu \nu}-\frac{1}{2}Rg_{\mu \nu}=8\pi G\Big((1+2\lambda)T_{\mu \nu}
+\lambda T g_{\mu \nu}+2\lambda \rho g_{\mu \nu}\Big) \,.
\end{align}
Accordingly, comparing to GR gravity,
the coefficient of metric on the right hand side
of equation (\ref{eq8}) might be considered as an effective
cosmological constant which depends on the trace $T$
\cite{frt11}.
On the other hand, the divergence of the energy-momentum tensor
according to field equations (\ref{eq8}), results in
\begin{align} \label{eq9}
\nabla_{\mu}T^{\mu}_{\nu}=-\frac{\lambda}{1+2\lambda}\partial_{\nu}\big(\rho+3p\big) \,.
\end{align}

Based on the cosmological principle, we further assume
that our universe is spatially homogeneous and isotropic on the
large scales. The metric of such universe is described by the
well-known Friedmann-Lema\^itre-Robertson-Walker (FLRW) spacetime.
For the spatially flat background, the corresponding line elements
takes the form
\begin{align} \label{eq10}
\mathrm{d}s^2=a^2(\tau)\big(-\mathrm{d}\tau^2+\mathrm{d}\vec{x}^2\big) \,.
\end{align}

Thereupon, modified Friedmann equations are given by
\begin{align}
& H^2=\frac{8\pi G}{3}\Big((1+\lambda)\sum_{i}\bar{\rho}_i-3\lambda \sum_{i}\bar{p}_i\Big) \,, \label{eq11} \\
& 2\frac{H'}{a}+3H^2=-8\pi G\Big(\lambda\sum_{i}\bar{\rho}_i+(1+5\lambda)\sum_{i}\bar{p}_i\Big) \,, \label{eq12}
\end{align}
where the prime indicates  deviation with respect to the conformal
time, $H=a'/a^2$ is the Hubble parameter, and index \textit{i}
indicates the component \textit{i}th in the universe filled with
radiation (R), baryons (B), dark matter (DM) and cosmological
constant ($\Lambda$). Taking account of equation (\ref{eq11}), for
the total density parameter
$\Omega_\mathrm{tot}=\sum_{i}\bar{\rho}_i/\rho_\mathrm{cr}$ (with
$\rho_\mathrm{cr}=3H^2/(8\pi G)$) in $f(R,T)$ gravity, we find
\begin{align} \label{eq13}
\Omega_\mathrm{tot}=\frac{1}{1+\lambda}\bigg(1+\frac{\lambda}{H^2}8\pi G\sum_{i}\bar{p}_i\bigg) \,.
\end{align}
Then, it is obvious that choosing $\lambda=0$ will restore
Einstein field equations in standard cosmology.

Notably, in the special case of a dust matter dominated
universe, equation (\ref{eq11}) reduces to $H^2=\frac{8\pi
G}{3}(1+\lambda)\sum_{i}\bar{\rho}_i$, which indicates that the
identified $f(R,T)$ model in equations (\ref{eq6}) and (\ref{eq7})
corresponds to a gravitational theory with a time dependent
cosmological constant, where the gravitational interaction between
matter and curvature is modified by the term $2f(T)$ in the
gravitational action, replacing the gravitational constant $G$ by
a running gravitational coupling parameter $G_{\rm eff}$
\cite{frt1,thermo1}.

Also it is important to study linear scalar perturbations
in modified $f(R,T)$ gravity, based on the perturbed
FLRW metric in the synchronous gauge (syn) given by
\begin{equation} \label{eq14}
\mathrm{d}s^2=a^2(\tau)\Big(-\mathrm{d}\tau^2+\big(\delta_{ij}+h_{ij}\big)\mathrm{d}x^i\mathrm{d}x^j\Big)
\end{equation}
where
\begin{align} \label{eq15}
h_{ij}(\vec{x},\tau)&=\int \mathrm{d}^3k\,e^{i\vec{k}.\vec{x}}
\bigg(\hat{k}_i\hat{k}_jh(\vec{k},\tau) \nonumber \\
&+\Big(\hat{k}_i\hat{k}_j-\frac{1}{3}\delta_{ij}\Big)6\eta(\vec{k},\tau)\bigg) \,,
\end{align}
with scalar perturbations $h$ and $\eta$,
and $\vec{k}=k\hat{k}$ \cite{pt}.
So, the corresponding modified field equations become
\begin{align}
\frac{a'}{a}h'-2k^2\eta&=8\pi G a^2 \Big((1+\lambda)\sum_{i}\delta \rho_{i(\mathrm{syn})} \nonumber \\
&-3\lambda\sum_{i}\delta p_{i(\mathrm{syn})}\Big) \,, \label{eq16}
\end{align}
\begin{align}
k^2\eta'=4\pi G(1+2\lambda) a^2 \sum_{i}\big(\bar{\rho}_i+\bar{p}_i\big)\theta_{i(\mathrm{syn})} \,, \label{eq17}
\end{align}
\begin{align}
\frac{1}{2}h''+3\eta''+\big(h'+6\eta'\big)\frac{a'}{a}-k^2\eta=0 \,, \label{eq18}
\end{align}
\begin{align}
-2\frac{a'}{a}h'-h''+2k^2\eta&=24\pi G a^2 \Big(\lambda\sum_{i}\delta \rho_{i(\mathrm{syn})} \nonumber \\
&+(1+5\lambda)\sum_{i}\delta p_{i(\mathrm{syn})}\Big)  \,. \label{eq19}
\end{align}
In parallel, for the perturbed FLRW metric in conformal
Newtonian gauge (con) described as
\begin{equation} \label{eq20}
\mathrm{d}s^2=a^2(\tau)\Big(-\big(1+2\Psi\big)\mathrm{d}\tau^2+\big(1-2\Phi\big)\mathrm{d}\vec{x}^2\Big) \,,
\end{equation}
with gravitational potentials $\Psi$ and $\Phi$ \cite{pt},
we find
\begin{align}
&k^2\Phi+3\frac{a'}{a}\Phi'+3\Big(\frac{a'}{a}\Big)^2\Psi \nonumber \\
&=4\pi G a^2 \Big(-(1+\lambda)\sum_{i}\delta \rho_{i(\mathrm{con})}+3\lambda\sum_{i}\delta p_{i(\mathrm{con})}\Big) \,, \label{eq21}
\end{align}
\begin{align}
& k^2\Phi'+\frac{a'}{a}k^2\Psi=4\pi G (1+2\lambda) a^2 \sum_{i}\big(\bar{\rho}_i+\bar{p}_i\big)\theta_{i(\mathrm{con})} \,, \label{eq22}
\end{align}
\begin{align}
\Phi-\Psi=0 \,, \label{eq23}
\end{align}
\begin{align}
&\bigg(2\frac{a''}{a}-\Big(\frac{a'}{a}\Big)^2\bigg)\Psi+\frac{a'}{a}\big(\Psi'+2\Phi'\big)+\Phi''+\frac{1}{3}k^2\big(\Phi-\Psi\big) \nonumber \\
&=4\pi G a^2 \Big(\lambda\sum_{i}\delta \rho_{i(\mathrm{con})}+(1+5\lambda)\sum_{i}\delta p_{i(\mathrm{con})}\Big) \,. \label{eq24}
\end{align}
Moreover, according to the non-conservation equation (\ref{eq9}),
one can rewrite the continuity equation in background level as
\begin{align}
\bar{\rho}'_i+\frac{3(1+w_i)(1+2\lambda)}{1+\lambda(1-3w_i)}\frac{a'}{a}\bar{\rho}_i=0 \,, \label{eq25}
\end{align}
in which we have considered $p_i=w_i\rho_i$ with constant
equation of state $w_i$. Correspondingly, to linear order of
perturbations (in synchronous gauge) we find
\begin{align}
\delta'_{i(\mathrm{syn})}=&\frac{1+2\lambda}{-1+\lambda(-1+3c^2_{si})} \nonumber \\
&\times \Bigg\{\delta_{i(\mathrm{syn})}\frac{a'}{a}\bigg(\frac{3(1+w_i)\big(-1+\lambda(-1+3c^2_{si})\big)}{1+\lambda(1-3w_i)} \nonumber \\
&+3(1+c^2_{si})\bigg) \nonumber \\
&+\frac{1}{2}h'(1+w_i)+(1+w_i)\theta_{i(\mathrm{syn})} \nonumber \\
&\times \bigg[1+9\frac{(c^2_{si}-c^2_{ai})(1+2\lambda)}{k^2\big(1+\lambda(1-3w_i)\big)} \nonumber \\
&\times \bigg(\Big(\frac{a'}{a}\Big)^2
-\frac{\lambda}{1+2\lambda}\Big(\frac{a''}{a} \nonumber \\
&-\Big(\frac{a'}{a}\Big)^2\Big(1+\frac{3(1+w_i)(1+2\lambda)}{1+\lambda(1-3w_i)}\Big)\Big)\bigg)\bigg] \nonumber \\
&-\frac{9(c^2_{si}-c^2_{ai})(1+w_i)\lambda}{k^2\big(1+\lambda(1-3w_i)\big)}\frac{a'}{a}\theta'_{i(\mathrm{syn})} \Bigg\} \,, \label{eq26}
\end{align}
\begin{align}
\theta'_{i(\mathrm{syn})}&=\theta_{i(\mathrm{syn})}\frac{a'}{a} \nonumber \\
&\times \bigg[\frac{3(1+w_i)(1+2\lambda)+3(1+5\lambda)(c^2_{si}-c^2_{ai})}{1+\lambda(1-3w_i)}-4\bigg] \nonumber \\
&+\frac{k^2}{(1+w_i)(1+2\lambda)}\Big(c^2_{si}+\lambda(1+5c^2_{si})\Big)\delta_{i(\mathrm{syn})} \,. \label{eq27}
\end{align}
Then, it can be easily understood that in case of $\lambda=0$,
the standard model of cosmology will be recovered.
Note that in the rest of the paper, we will study the
$f(R,T)$ model in the synchronous gauge.

Now that we have derived modified equations in $f(R,T)$ gravity,
we are in the position to investigate the cosmological properties
of $f(R,T)$ model by the publicly available Boltzmann code CLASS
\footnote{Cosmic Linear Anisotropy Solving System} \cite{cl},
as well as utilizing the MCMC\footnote{Markov Chain Monte Carlo}
package M\textsc{onte} P\textsc{ython} \cite{mp1,mp2} in order
to compare the model with observations.
\section{Cosmological results} \label{sec3}
In the purpose of exploring the cosmological observables, mainly
CMB anisotropy and matter power spectra, in $f(R,T)$ model, we
modify the CLASS code to incorporate the model parameter $\lambda$
together with derived equations described in sections \ref{sec2}.
In this approach, the cosmological parameters are considered based
on Planck 2018 data \cite{cmb3}, given by
$\Omega_{\mathrm{B},0}h^2=0.02242$,
$\Omega_{\mathrm{DM},0}h^2=0.11933$,
$H_0=67.66\,\mathrm{\frac{km}{s\,Mpc}}$, $A_s=2.105\times
10^{-9}$, and $\tau_\mathrm{reio}=0.0561$.

\begin{figure*}[ht!]
    \includegraphics[width=8.5cm]{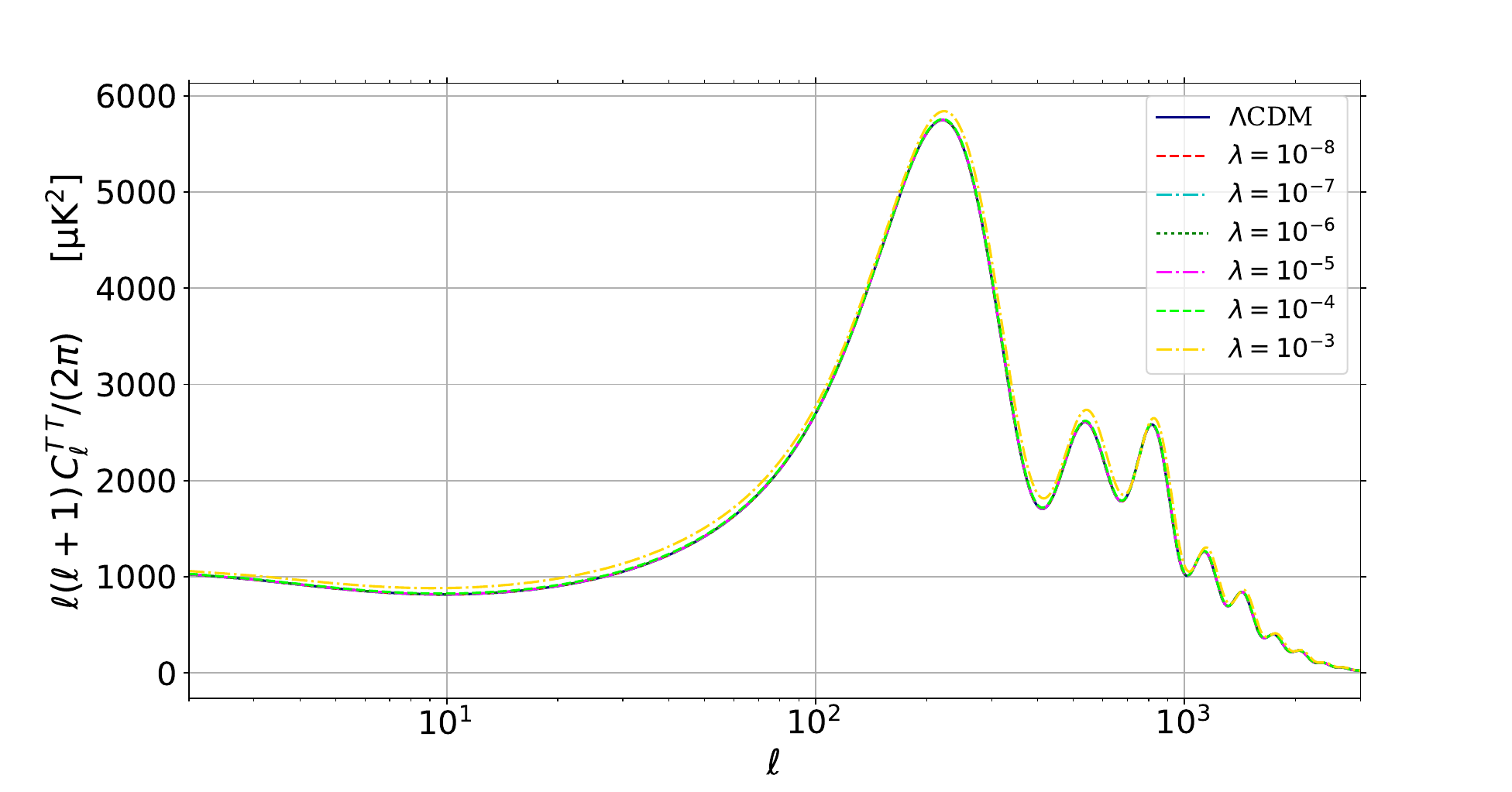}
    \includegraphics[width=8.5cm]{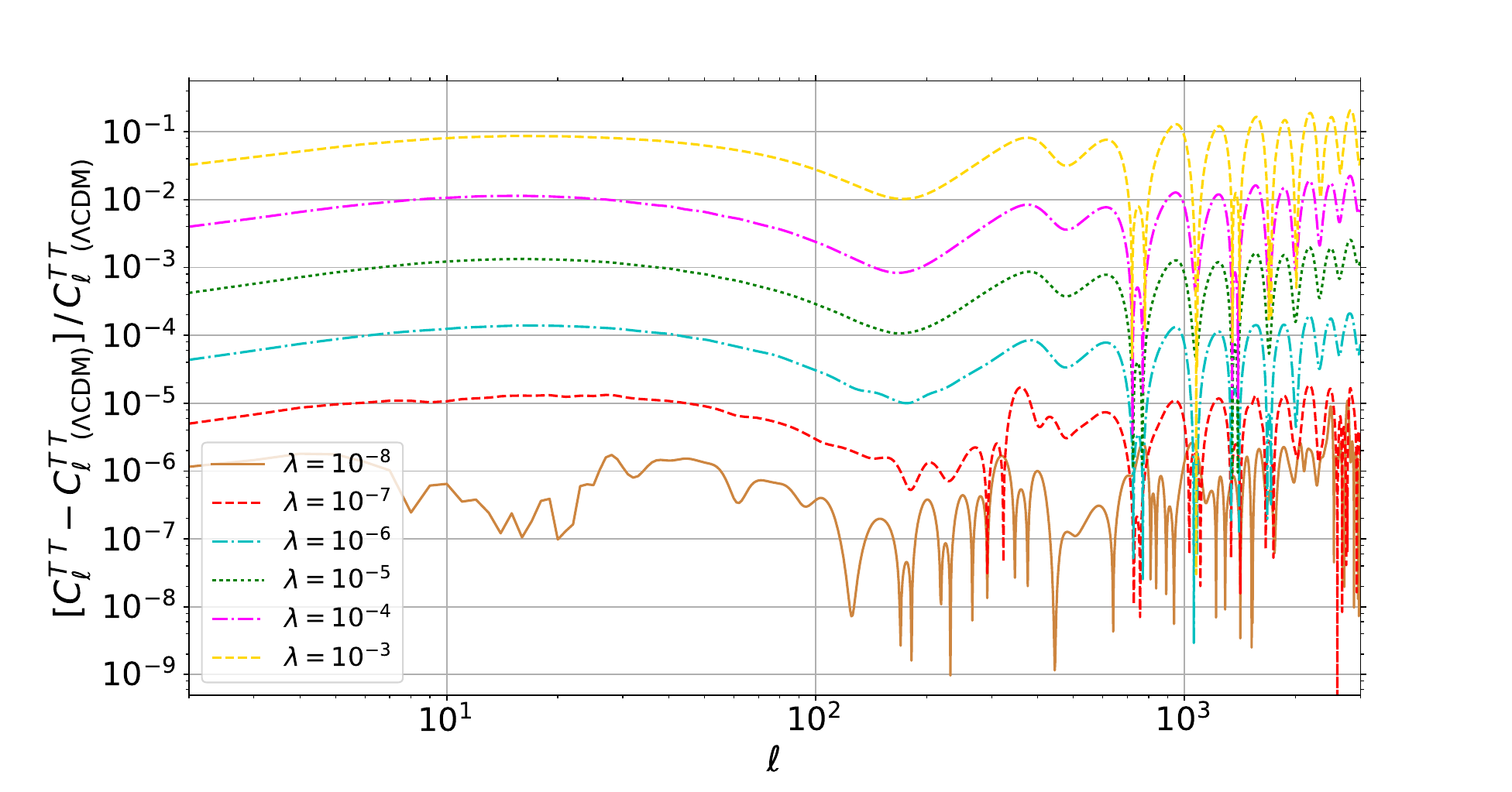}
    \caption{The CMB power spectra diagrams (left panel)
        and their relative ratio with respect to the $\Lambda$CDM
        model (right panel) for different values of $\lambda$.}
    \label{fig1}
\end{figure*}

\begin{figure*}[ht!]
    \includegraphics[width=8.5cm]{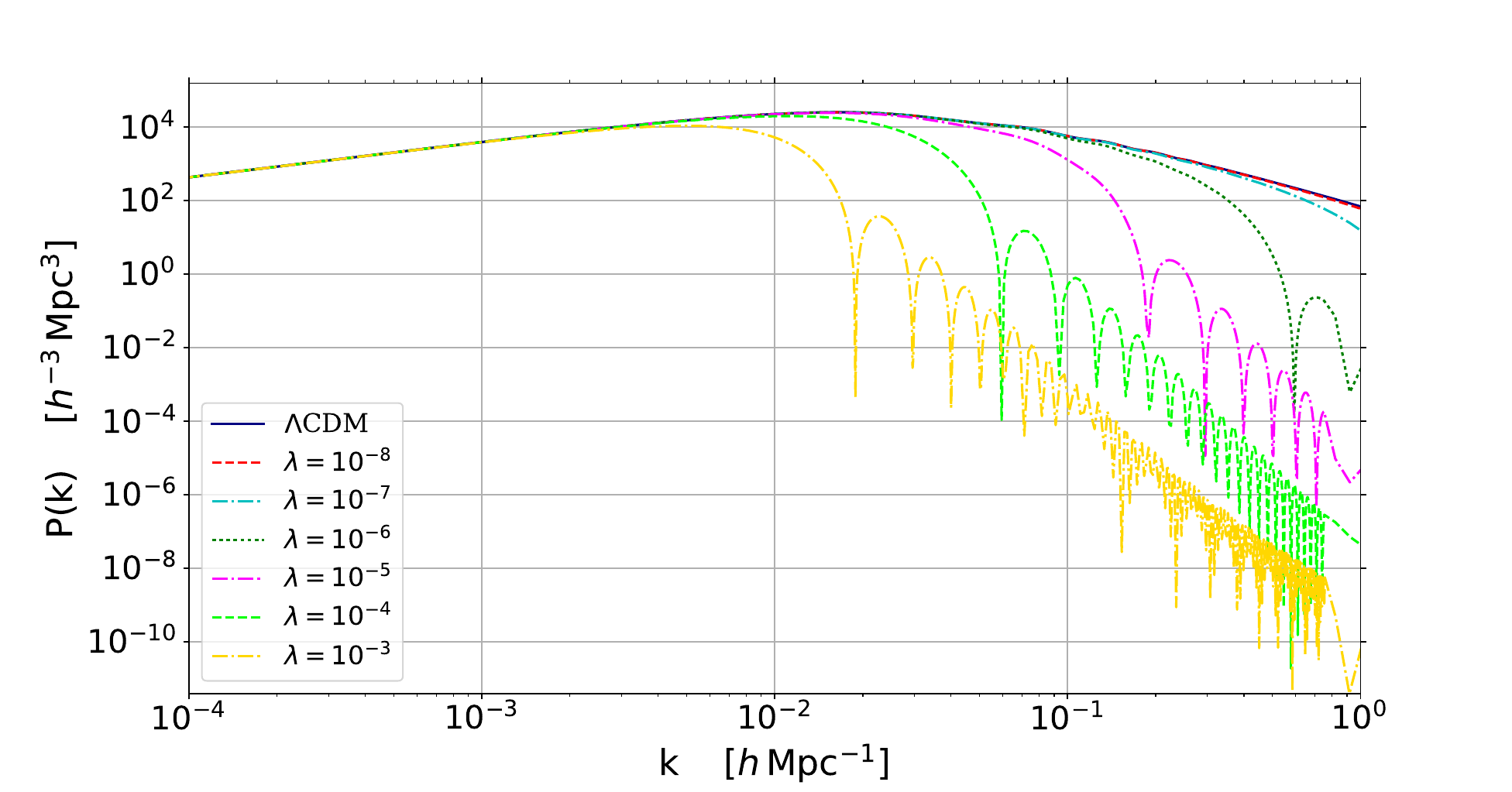}
    \includegraphics[width=8.5cm]{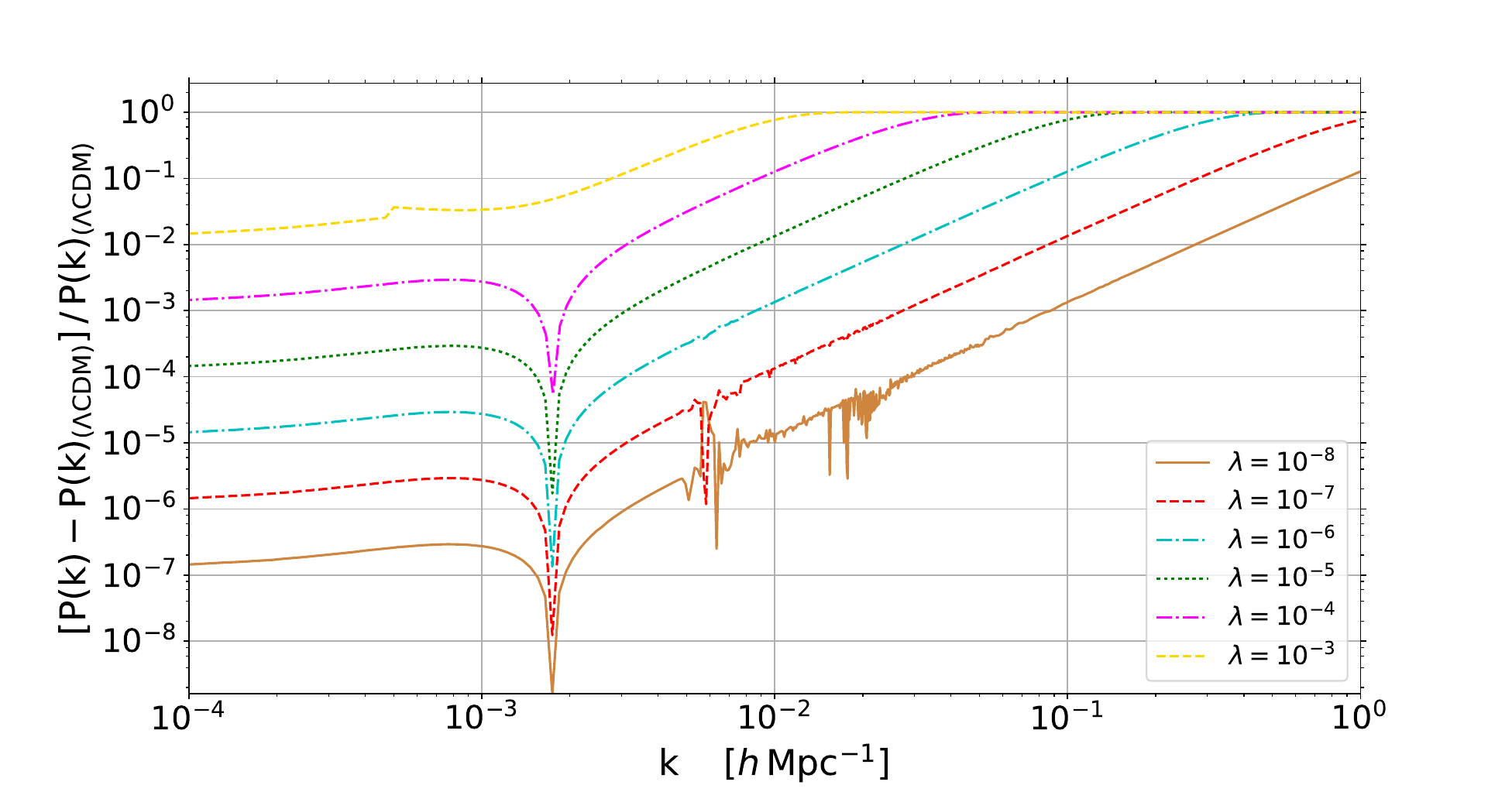}
    \caption{The matter power spectra diagrams (left panel) and
        their relative ratio with respect to the $\Lambda$CDM model
        (right panel) for different values of $\lambda$.}
    \label{fig2}
\end{figure*}

\begin{figure*}[ht!]
    \includegraphics[width=8.5cm]{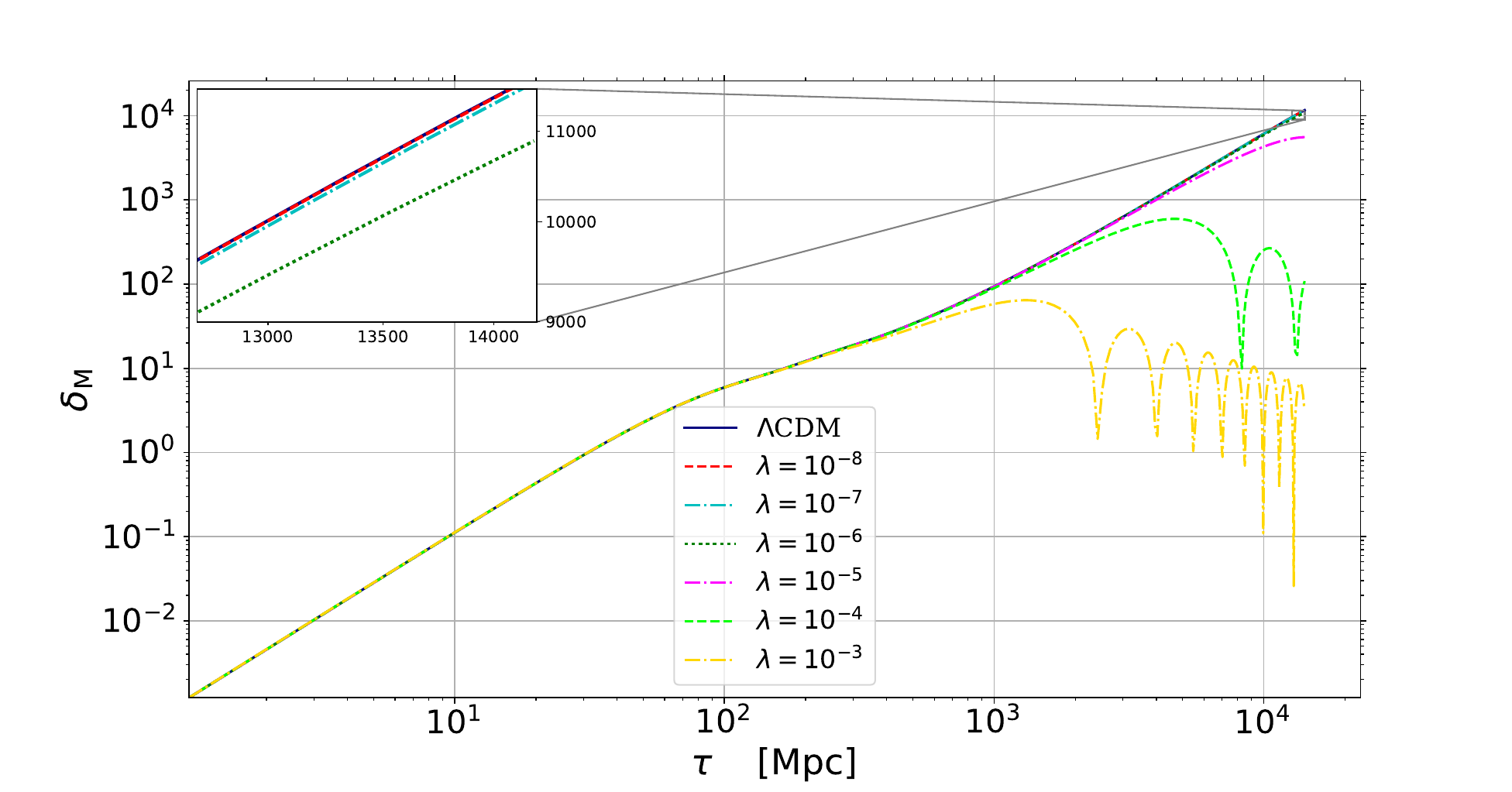}
    \includegraphics[width=8.5cm]{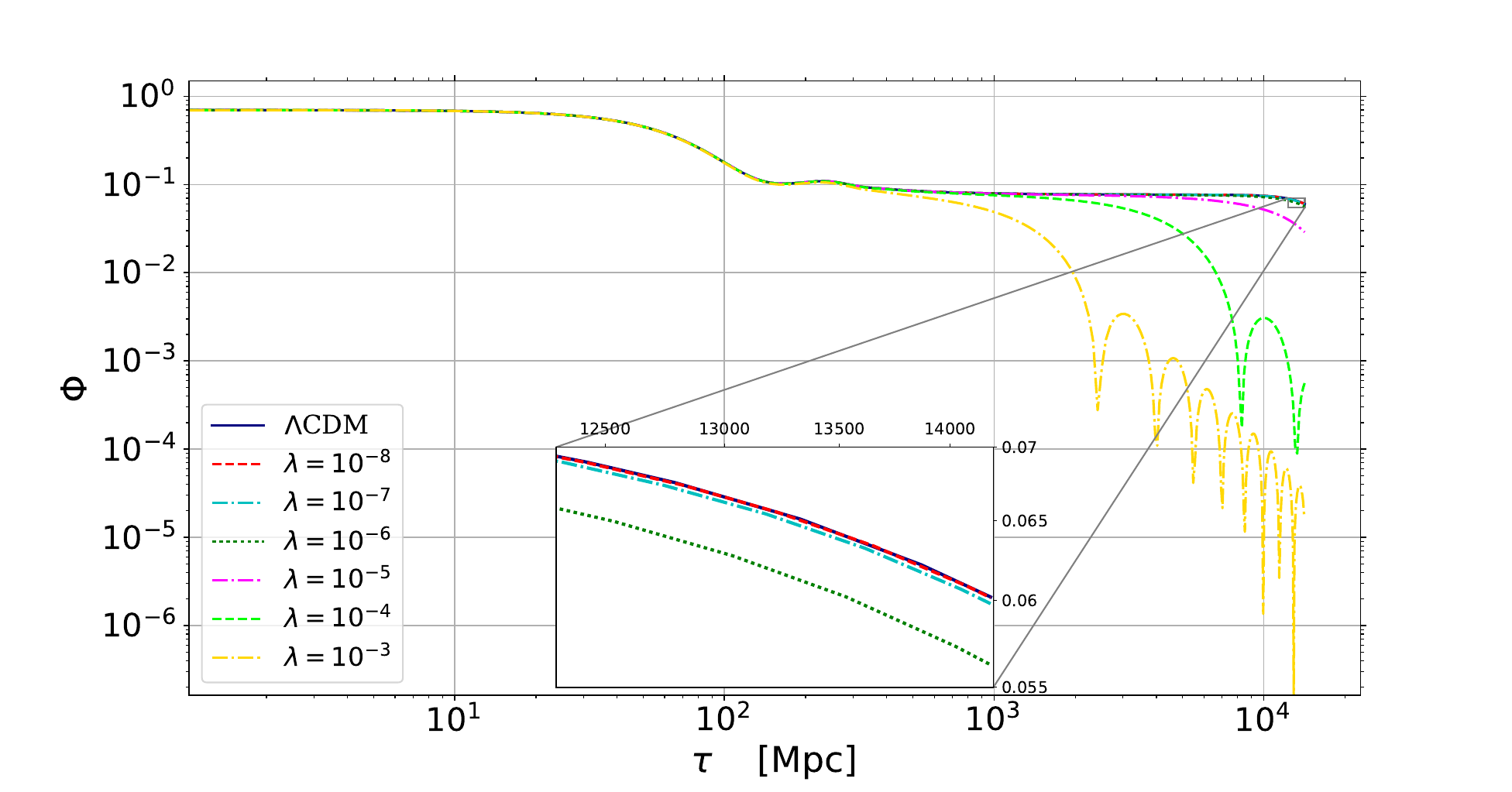}
    \caption{The matter density contrast diagrams (left panel)
        and the Newtonian potential diagrams (right panel) for
        different values of $\lambda$, compared to $\Lambda$CDM.}
    \label{fig3}
\end{figure*}

In figure (\ref{fig1}) we present the CMB anisotropy power
spectra in $f(R,T)$ gravity with different values of $\lambda$,
where for more clarification their relative ratio with respect
to the concordance $\Lambda$CDM model are also displayed.
More interestingly, in sake of probing structure formation
in modified $f(R,T)$ gravity, one can consider matter power
spectra diagrams as depicted in figure (\ref{fig2}).
According to this figure, we realize a suppression in the
structure growth in $f(R,T)$ model which reports consistency
with local observations \cite{s81,s82,s83,s84,s85,s86}.
In addition, figure (\ref{fig3}) illustrates the matter density
perturbations (left panel) and Newtonian potential (right panel),
which confirm the reduction in the growth of structures considering
$f(R,T)$ modified theory of gravity.

\begin{figure*}[ht!]
    \includegraphics[width=8.5cm]{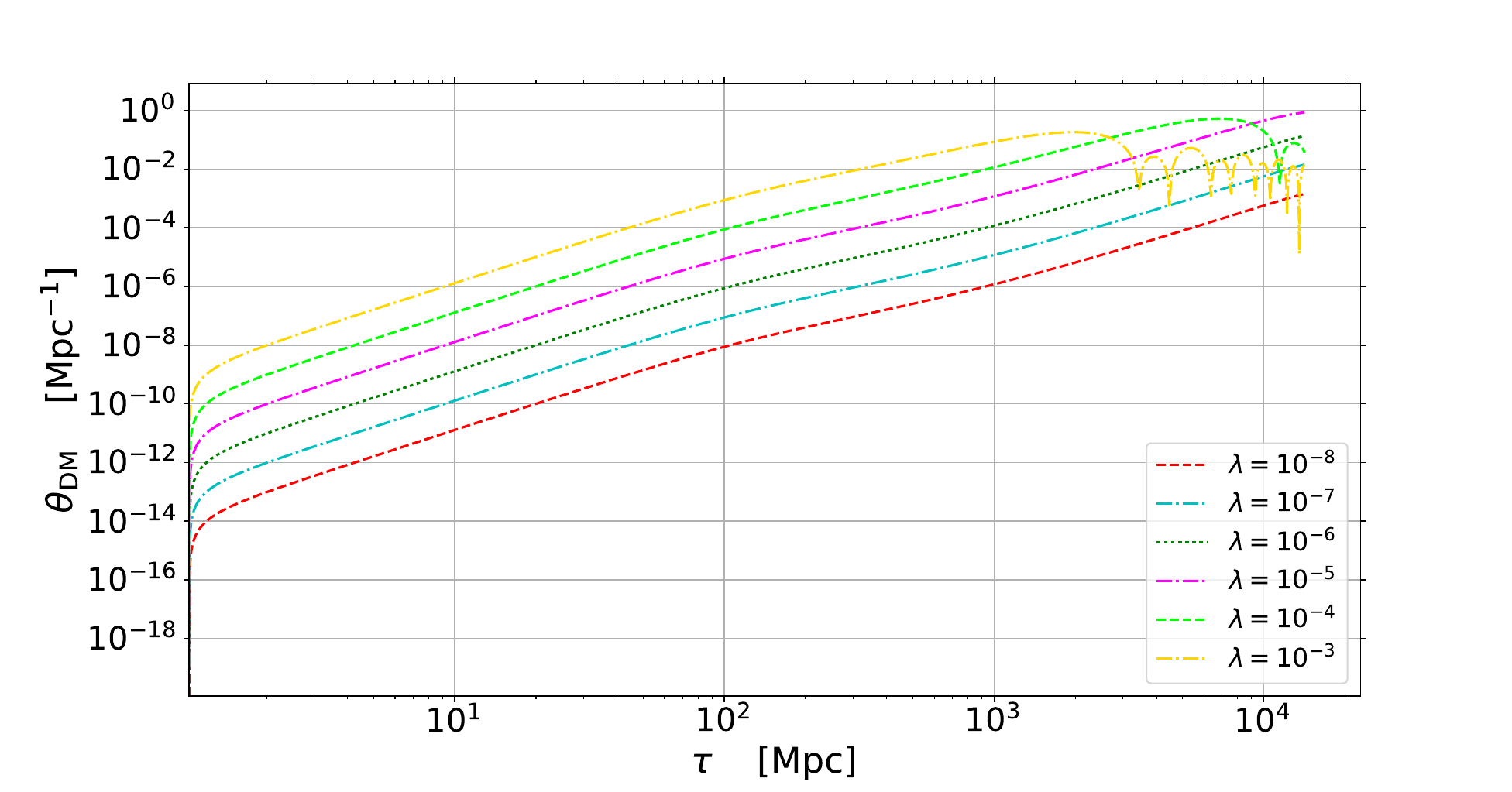}
    \includegraphics[width=8.5cm]{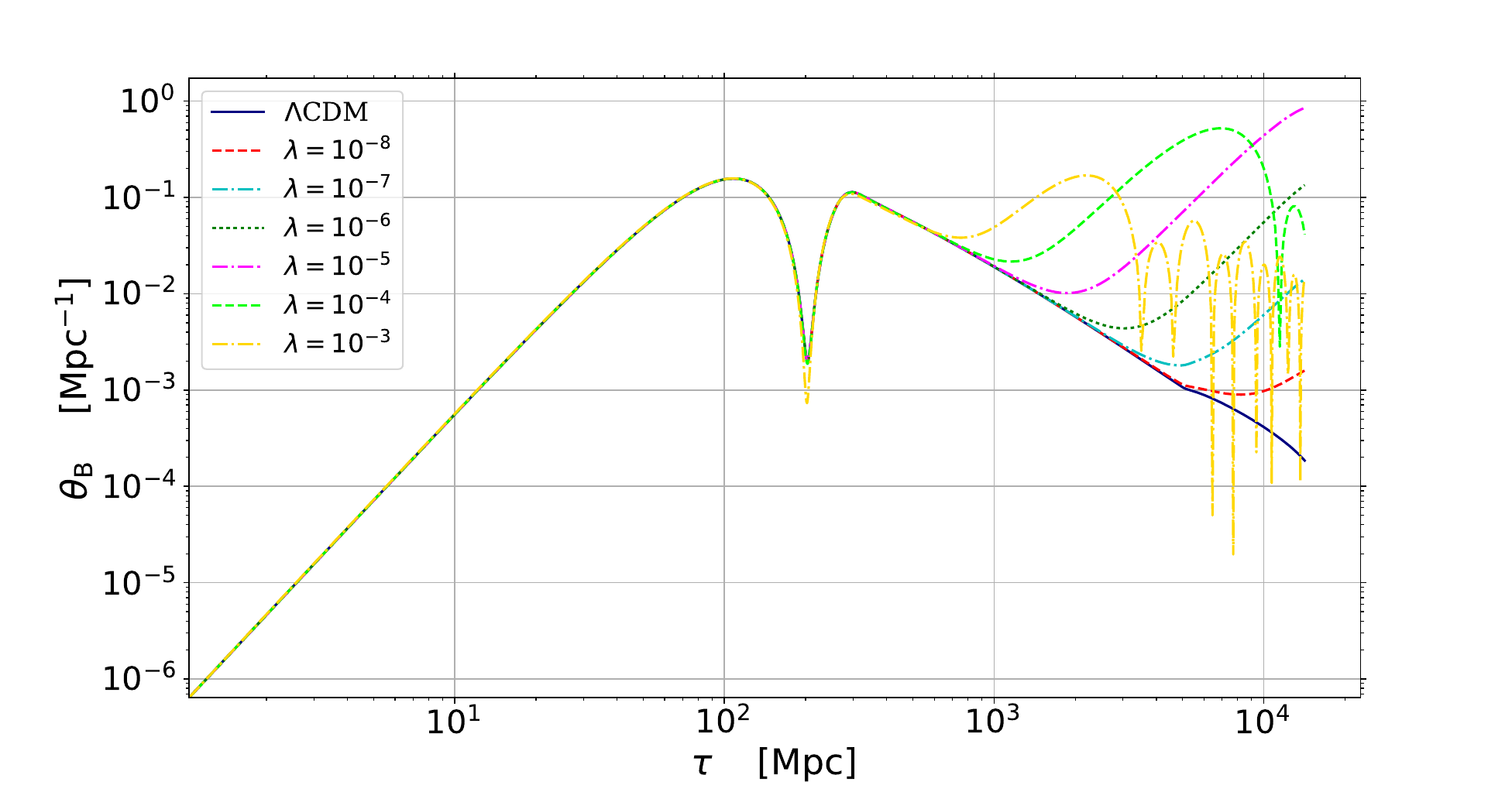}
    \caption{The velocity perturbations of dark matter
        (left panel) and baryons (right panel) for different values
        of $\lambda$.}
    \label{fig4}
\end{figure*}

Also, an intriguing feature is evident from matter power
spectra as well as matter density and potential perturbation
diagrams, contemplating as "matter acoustic oscillations" in
$f(R,T)$ model, due to non-conservation equation at perturbation
level. These oscillations can be induced by the particle creation
process as a result of matter-geometry interaction. According to
a thermodynamic point of view, irreversible matter creation
processes produce negative pressure \cite{thermo1,thermo2},
which seems to be effective in suppressing structure formation.
Matter acoustic oscillations are more apparent in velocity
perturbations shown in figure (\ref{fig4}).

\begin{figure}[ht!]
    \centering
    \includegraphics[width=8.5cm]{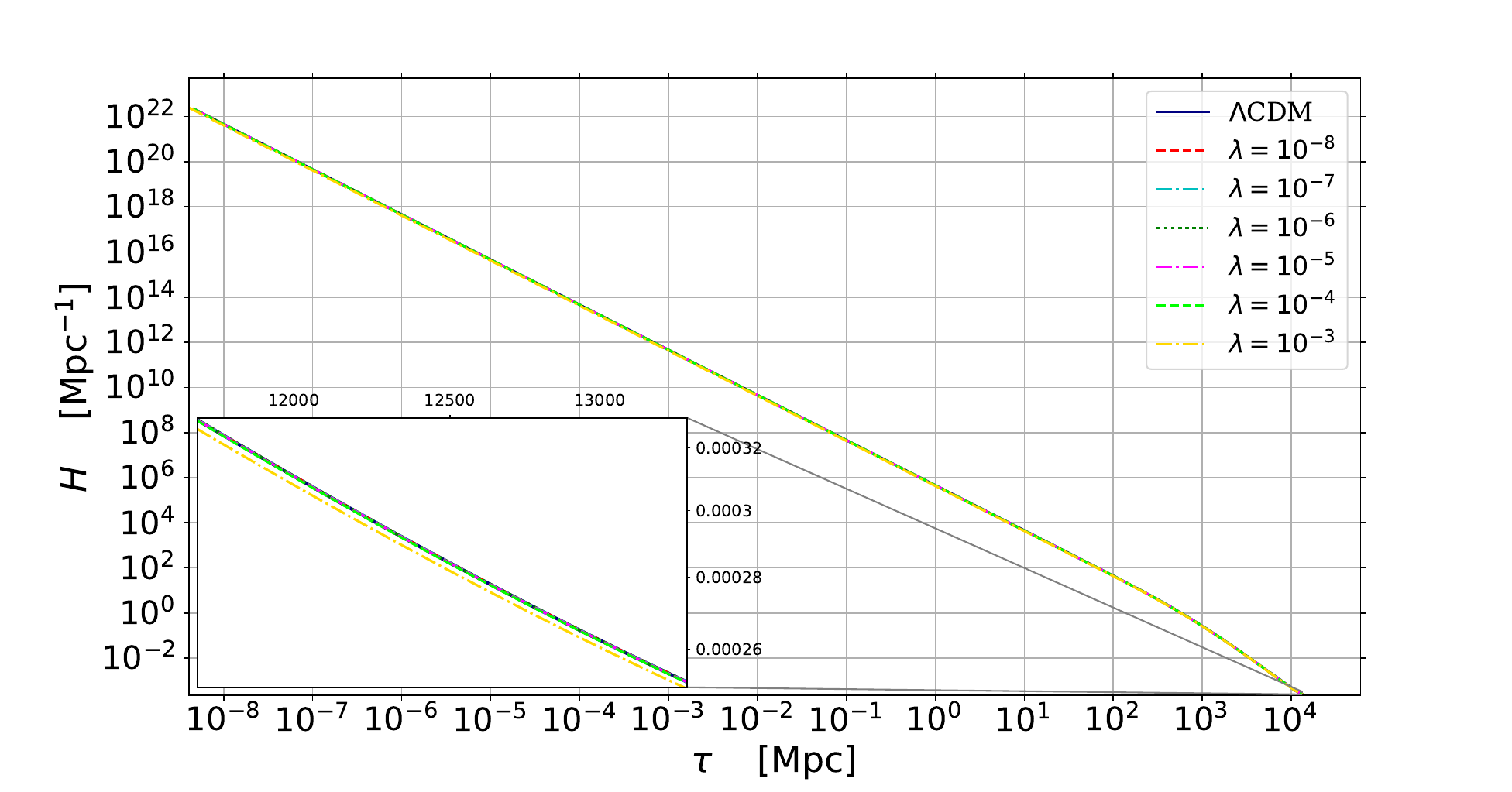}
    \caption{Hubble parameter for different values of $\lambda$,
        compared to $\Lambda$CDM.}
    \label{fig5}
\end{figure}

For the last point, we turn our attention to the expansion history of
the universe in $f(R,T)$ gravity. To this aim, the Hubble parameter
diagrams in $f(R,T)$ model compared to $\Lambda$CDM are displayed
in figure (\ref{fig5}). Accordingly, we detect lower values of $H_0$
in $f(R,T)$ theory, which means that Hubble tension might become
more severe in this model of modified gravity.
\section{Constraints on cosmological parameters} \label{sec4}
Having studied the cosmological features of modified $f(R,T)$
gravity, we proceed to probe observational constraints on the
model parameters, as well as investigating the capability of
$f(R,T)$ model in ameliorating the $\sigma_8$ tension.
In this respect, we perform an MCMC approach through the
publicly available code M\textsc{onte} P\textsc{ython}.
The set of cosmological parameters employed in numerical
calculation is \{ $100\,\Omega_{\mathrm{B},0} h^2$,
$\Omega_{\mathrm{DM},0} h^2$, $100\,\theta_s$, $\ln (10^{10}
A_s)$, $n_s$, $\tau_{\mathrm{reio}}$, $\lambda$ \},
consists of the six cosmological parameters in standard
$\Lambda$CDM model, along with the $f(R,T)$ model parameter
$\lambda$, which quantifies deviations from standard cosmology.
One should be noted that primary numerical investigations
suggest the prior range [$0$, $10^{-4}$] for $\lambda$.
We have also four derived parameters including the
reionization redshift $z_\mathrm{reio}$, the matter
density parameter $\Omega_{\mathrm{M},0}$, the Hubble constant
$H_0$, and the structure growth parameter $\sigma_8$.

The specific likelihoods utilized in MCMC technique contains
the Planck likelihood with Planck 2018 data including high-$l$
TT,TE,EE, low-$l$ EE, low-$l$ TT, and lensing measurements
\cite{cmb3}, the Planck-SZ likelihood for the Sunyaev-Zeldovich
effect measured by Planck \cite{sz1,sz2},
the CFHTLenS likelihood with the weak lensing data
\cite{lens1,lens2}, the Pantheon likelihood with the supernovae
data \cite{pan},  the BAO likelihood with the baryon acoustic
oscillations data \cite{bao4,bao5}, and the BAORSD likelihood for
BAO and redshift-space distortions measurements
\cite{rsd1,rsd2}.

\begin{figure}[h!]
    \centering
    \includegraphics[width=8.5cm]{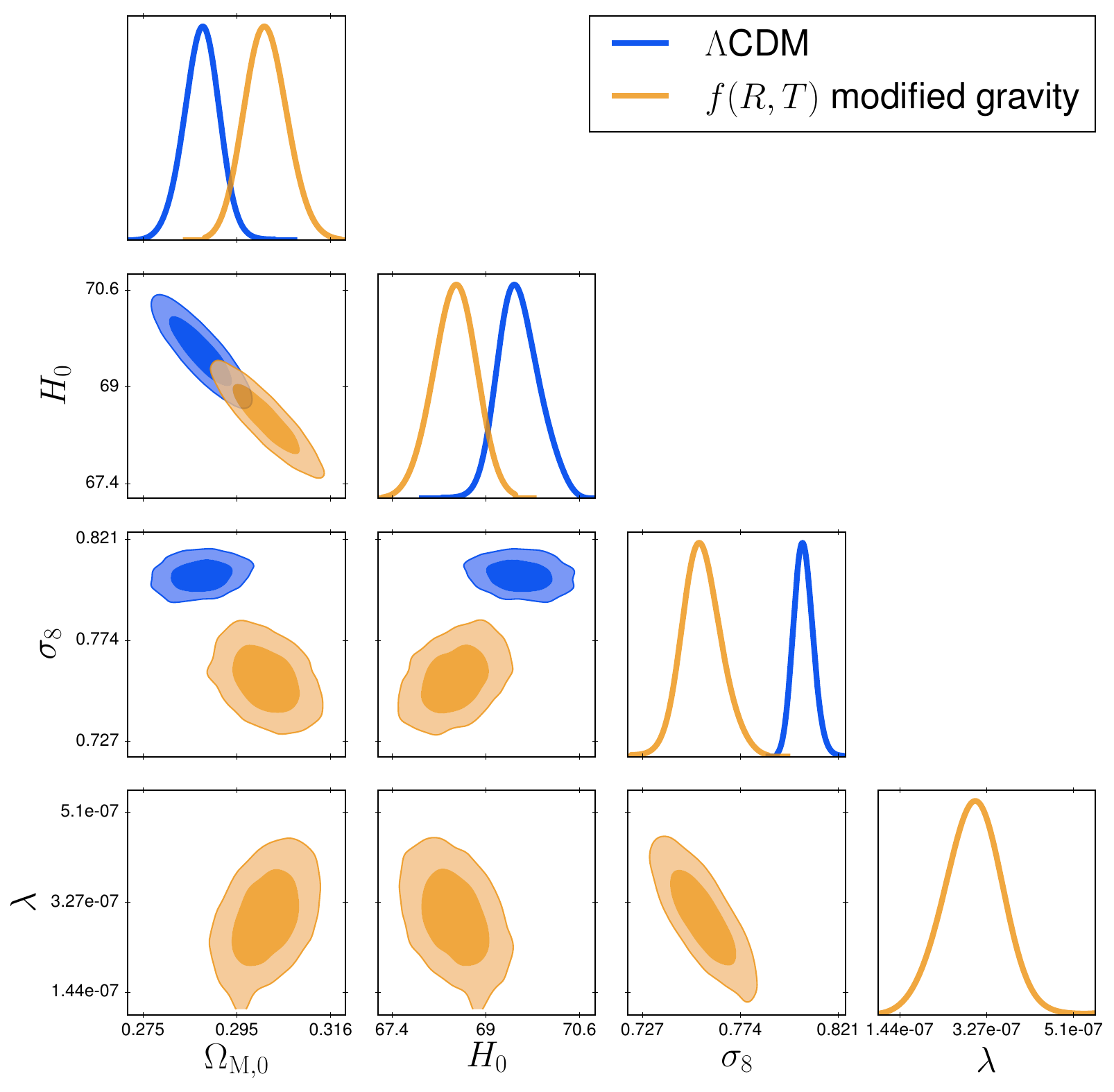}
    \caption{The $1\sigma$ and $2\sigma$ constraints from the
        "Planck + Planck-SZ + CFHTLenS + Pantheon + BAO + BAORSD"
        dataset on some selected cosmological parameters of $f(R,T)$
        gravity compared to $\Lambda$CDM model.}
    \label{fig6}
\end{figure}

Table \ref{tab1} reports the derived observational constraints on
cosmological parameters based on the combined dataset
"Planck + Planck-SZ + CFHTLenS + Pantheon + BAO + BAORSD".
Furthermore, we demonstrate the one-dimensional posterior
probabilities and two-dimensional contours for some selected
cosmological parameters of $f(R,T)$ gravity compared to standard
cosmological model in figure (\ref{fig6}).
\begin{table}
    \centering
    \caption{Best fit values of cosmological parameters with
        the $1\sigma$ and $2\sigma$ confidence levels using
        "Planck + Planck-SZ + CFHTLenS + Pantheon + BAO + BAORSD"
        dataset for $\Lambda$CDM model and $f(R,T)$ gravity.}
    \scalebox{.7}{
        \begin{tabular}{|c|c|c|c|c|}
            \hline
            & \multicolumn{2}{|c|}{} & \multicolumn{2}{|c|}{} \\
            & \multicolumn{2}{|c|}{$\Lambda$CDM} & \multicolumn{2}{|c|}{$f(R,T)$ gravity} \\
            \cline{2-5}
            & & & & \\
            {parameter} & best fit & $1\sigma$ \& $2\sigma$ limits & best fit & $1\sigma$ \& $2\sigma$ limits \\ \hline
            & & & & \\
            $100\,\Omega_{\mathrm{B},0} h^2$ & $2.261$ & $2.263^{+0.012+0.026}_{-0.013-0.025}$ & $2.249$ & $2.246^{+0.013+0.027}_{-0.013-0.026}$ \\
            & & & & \\
            $\Omega_{\mathrm{DM},0} h^2$ & $0.1163$ & $0.1164^{+0.00078+0.0015}_{-0.00079-0.0015}$ & $0.1190$ & $0.1189^{+0.00082+0.0017}_{-0.00081-0.0017}$ \\
            & & & & \\
            $100\,\theta_s$ & $1.042$ & $1.042^{+0.00029+0.00055}_{-0.00026-0.00053}$ & $1.042$ & $1.042^{+0.00027+0.00058}_{-0.00030-0.00060}$ \\
            & & & & \\
            $\ln (10^{10} A_s)$ & $3.034$ & $3.024^{+0.010+0.023}_{-0.014-0.021}$ & $3.052$ & $3.050^{+0.014+0.032}_{-0.016-0.028}$ \\
            & & & & \\
            $n_s$ & $0.9712$ & $0.9719^{+0.0036+0.0072}_{-0.0039-0.0074}$ & $0.9664$ & $0.9682^{+0.0035+0.0070}_{-0.0036-0.0068}$ \\
            & & & & \\
            $\tau_\mathrm{reio}$ & $0.05358$ & $0.04963^{+0.0041+0.010}_{-0.0074-0.0096}$ & $0.05939$ & $0.05772^{+0.0067+0.015}_{-0.0081-0.015}$ \\
            & & & & \\
            $\lambda$ & --- & --- & $2.682\mathrm{e}{-7}$ & $2.972\mathrm{e}{-7}^{+6.0\mathrm{e}{-8}+1.3\mathrm{e}{-7}}_{-6.5\mathrm{e}{-8}-1.2\mathrm{e}{-7}}$ \\
            & & & & \\
            $z_\mathrm{reio}$ & $7.502$ & $7.084^{+0.50+1.0}_{-0.69-1.0}$ & $8.152$ & $7.976^{+0.66+1.4}_{-0.79-1.5}$ \\
            & & & & \\
            $\Omega_{\mathrm{M},0}$ & $0.2871$ & $0.2876^{+0.0043+0.0086}_{-0.0044-0.0086}$ & $0.3022$ & $0.3021^{+0.0047+0.0095}_{-0.0050-0.0098}$ \\
            & & & & \\
            $H_0\;[\mathrm{\frac{km}{s\,Mpc}}]$ & $69.56$ & $69.54^{+0.37+0.73}_{-0.36-0.71}$ & $68.43$ & $68.42^{+0.37+0.76}_{-0.37-0.73}$ \\
            & & & & \\
            $\sigma_8$ & $0.8079$ & $0.8044^{+0.0045+0.0096}_{-0.0051-0.0091}$ & $0.7623$ & $0.7561^{+0.0096+0.021}_{-0.010-0.019}$ \\
            & & & & \\
            \hline
        \end{tabular}
    }
    \label{tab1}
\end{table}
Numerical results indicate that $f(R,T)$ gravity anticipates a
lower value for the structure growth parameter $\sigma_8$ compared
to the standard $\Lambda$CDM model. So, we perceive that $f(R,T)$
model is capable of alleviating the $\sigma_8$ tension, by
reconciling low redshift measurements of cosmic structure growth
with Planck data. Moreover, according to the obtained constraints
on Hubble constant, it seems that $H_0$ tension becomes more
serious in $f(R,T)$ model, in accordance with the correlation
between $H_0$ and $\sigma_8$.

On the other hand, obtained constraints on the $f(R,T)$ model
parameter $\lambda$, report deviations from $\Lambda$CDM at more
than $3\sigma$. Accordingly, in order to comprehend the
more compatible model with observations, we apply the Akaike
information criterion (AIC) defined as
\cite{aic1,aic2}
\begin{equation}
\mathrm{AIC}=-2\ln{\mathcal{L}_{\mathrm{max}}}+2K ,
\end{equation}
where $\mathcal{L}_{\mathrm{max}}$ is the maximum likelihood
function and $K$ stands for the number of free parameters.
Then, MCMC calculations result in $\mathrm{AIC_{(\Lambda CDM)}}=3847.12$,
and $\mathrm{AIC_{(modified\;gravity)}}=3829.82$, which conclude
$\mathrm{\Delta AIC}=17.3$.
Hence, we detect a strong evidence in favor of $f(R,T)$ gravity
which means that the studied $f(R,T)$ model is worth further
precise numerical investigations with a variety of reliable
observational data.
\section{Conclusions} \label{sec5}
In the present investigation, we have studied the compatibility
of $f(R,T)$ modified gravity with observations, as well as its
potentiality in reconciling the cosmological tensions.
$f(R,T)$ model in which the gravitational Lagrangian is
considered as a function of the Ricci scalar $R$ and the trace
of the energy-momentum tensor $T$, is a generalized modified
gravity based on a non-minimal matter-geometry interaction
\cite{frt1}. We specify the functional expression described
in equations (\ref{eq6}) and (\ref{eq7}) for $f(R,T)$,
which yields modified field equations derived in section
\ref{sec2}. Then, by employing the modified version of the
CLASS code based on $f(R,T)$ model, we are in the position to
explore the cosmological observables in the framework of
$f(R,T)$ gravity.
Accordingly, matter power spectra diagrams
represented in figure (\ref{fig2}), report a cosmic
structure growth suppression in $f(R,T)$ model,
which is compatible with low redshift structure growth
determinations. Furthermore, exploring the evolution of
matter density contrast as well as the Newtonian potential,
verifies the suppressed structure growth in $f(R,T)$ gravity
as displayed in figure (\ref{fig3}).
Moreover, we detect the remarkable feature
"matter acoustic oscillations" in $f(R,T)$ gravity due to the
coupling between matter and geometry.
The velocity perturbations diagrams illustrated in
figure (\ref{fig4}) exhibit matter acoustic oscillations
more distinctly. The matter-curvature interaction
produces particles, where this irreversible
matter creation process generates negative pressure.
Consequently, the negative pressure is effectual in
the suppression of structure growth \cite{thermo1,thermo2}.
Thus, primary numerical studies imply that modified $f(R,T)$
gravity proves beneficial in relieving the observed
tension between local and global probes of structure formation.
On the other hand, Hubble parameter diagrams
in figure (\ref{fig5}) reveals that the Hubble tension becomes
more serious in $f(R,T)$ gravity.
Correspondingly, in pursuance of
examining the qualification of $f(R,T)$ model to reduce the
cosmological tensions, we constrain the model with current
observational data. For this purpose, we consider Planck CMB,
weak lensing, supernovae, BAO, and RSD measurements in our
MCMC analysis. According to MCMC numerical investigations,
one can conclude that $f(R,T)$ modified gravity represents
capability in ameliorating the $\sigma_8$ tension, by
predicting lower structure growth for the universe.
It is also notable that, the Hubble tension
becomes more severe in $f(R,T)$ gravity, and thus,
a simultaneous alleviation of both the $H_0$ and $\sigma_8$
tensions is not possible in the studied $f(R,T)$ model.
\section*{Declaration of competing interest}
The authors declare that they have no known competing financial interests
or personal relationships that could have appeared to influence the work
reported in this paper.
\section*{Data availability}
No new data were generated or analysed during the current study.
\section*{Acknowledgments}
We thank Shiraz University Research Council.
This work is based upon research funded by
Iran National Science Foundation (INSF) under project No. 4024184.
We are also grateful to the referee for valuable comments
which helped us improve the paper significantly.

\interlinepenalty=10000
\bibliography{1}

\end{document}